# Soil Property and Class Maps of the Conterminous United States at 100-Meter Spatial Resolution

**Amanda Ramcharan***
Penn State Univ.
University Park, PA 16802

**Tomislav Hengl**
ISRIC—World Soil Information
Wageningen, The Netherlands

**Travis Nauman**
USGS
Southwest Biological Science Center
Moab, UT 84532

**Colby Brungard**
New Mexico State Univ.
Las Cruces, NM 88003

**Sharon Waltman**
West Virginia Univ.— GRU
Morgantown, WV 26506

**Skye Wills**
NRCS
Lincoln, NE 68508

**James Thompson**
West Virginia Univ.
Morgantown, WV 26506

With growing concern for the depletion of soil resources, conventional soil maps need to be updated and provided at finer and finer resolutions to be able to support spatially explicit human–landscape models. Three US soil point datasets—the National Cooperative Soil Survey Characterization Database, the National Soil Information System, and the Rapid Carbon Assessment dataset—were combined with a stack of over 200 environmental datasets and gSSURGO polygon maps to generate complete coverage gridded predictions at 100-m spatial resolution of six soil properties (percentage of organic C, total N, bulk density, pH, and percentage of sand and clay) and two US soil taxonomic classes (291 great groups [GGs] and 78 modified particle size classes [mPSCs]) for the conterminous United States. Models were built using parallelized random forest and gradient boosting algorithms as implemented in the ranger and xgboost packages for R. Soil property predictions were generated at seven standard soil depths (0, 5, 15, 30, 60, 100, and 200 cm). Prediction probability maps for US soil taxonomic classifications were also generated. Cross validation results indicated an out-of-bag classification accuracy of 60% for GGs and 66% for mPSCs; for soil properties, RMSE for leave-location-out cross-validation was 0.74 ($R^2$ = 0.68), 17.8 wt% ($R^2$ = 0.57), 12 wt% ($R^2$ = 0.46), 3.63 wt% ($R^2$ = 0.41), 0.2 g cm$^{-3}$ ($R^2$ = 0.42), and 0.27 wt% ($R^2$ = 0.39) for pH, percent sand and clay, weight percentage of organic C, bulk density, and weight percentage of total N, respectively. Nine independent validation datasets were used to assess prediction accuracies for soil class models, and results ranged between 24 and 58% and between 24 and 93% for GG and mPSC prediction accuracies, respectively. Although mapping accuracies were variable and likely lower than gSSURGO in some areas, this modeling approach can enable easier integration of soil information with spatially explicit models compared with multicomponent map units.

**Abbreviations:** DEM, digital elevation model; GG, great group; mPSC, modified particle size class; NASIS, National Soil Information System; NCSS, National Cooperative Soil Survey; RaCA, Rapid Carbon Assessment; SOC, soil organic carbon; SSURGO, Soil Survey Geographic database; VIMP, variable importance.

T he National Academy of Sciences has reported that changes in the quality of soil and water resources have accelerated during the past decades at a rate proportional to human-induced activities and population growth (National Research Council, 2001). With the growing concern about finite soil resources, soil scientists are tasked with providing information that can support spatially explicit human–landscape models that can aid in preserving soil resources. This information needs to meet current requirements to estimate and manage stores and fluxes of water, carbon, nutrients, and solutes in soil. Conventional systems for mapping and classifying soils were not designed for this purpose (Arrouays et al.,

### Core Ideas

- Ensemble machine learning methods were used to obtain gridded soil property and class maps.
- Final predictions were generated for six soil properties and two soil classes at 100-m resolution.
- Soil data are easier to integrate with spatially explicit models compared with multicomponent map units.
- Soil property maps are available at seven standard depths.

2014). Instead, a gridded product with three-dimensional (3D) estimates of soil properties is required.

The highest-resolution 3D soil database with conterminous coverage of the United States is the vector Soil Survey Geographic database (SSURGO) and its 10-m resolution gridded equivalent gSSURGO (https://gdg.sc.egov.usda.gov/). Because this database compiles approximately 3000 independent soil surveys varying in scale from 1:12,000 to 1:125,000 (Thompson et al., 2012), associated tabular attributes are of different scales, and map units comprise different numbers of soil components. Within a map unit, soil components are not spatially assigned, and therefore spatial disaggregation is required to derive a gridded product at regional or national scales. Odgers et al. (2012) produced 100-m resolution soil property maps from the US General Soil Map (STATSGO2) by combining spline fitting and weighted averaging per mapping unit. Although nominally at high spatial resolution, maps produced by Odgers et al. (2012) reveal a relatively coarse scale of the STATSGO2 (1:250,000). Chaney et al. (2016) recently disaggregated and harmonized SSURGO components within the SSURGO database to create POLARIS, a 30-m spatial resolution map of the 50 most probable SSURGO components. Linking soil properties to POLARIS is complex due to low accuracies, and therefore it remains to be determined whether creating soil property maps from SSURGO component predictions is the optimal path to follow.

Acknowledging the complexities of creating gridded soil property maps from conventional soil maps, such as artificial boundaries in the data associated with geopolitical boundaries, unnamed soil components, and variation in soil properties in soil components without location of these components within the larger map unit delineation (Arrouays et al., 2014; Batjes, 2016; Odgers et al., 2012; Thompson et al., 2010), we have opted here for a point data/machine learning-driven 3D approach to predictive soil mapping. Machine learning algorithms, such as regression trees, Cubist, and random forests, have been proven useful for predictive soil mapping by Henderson et al. (2004) and Kheir et al. (2010) and then later by Hengl et al. (2014, 2017), who have developed a fully automated framework for global soil property and class mapping called "SoilGrids." One limitation of the (global) SoilGrids system is the inability to integrate detailed national data into the modeling system. Integrating national soil and environmental datasets may result in predictions with a higher accuracy than previous SoilGrids results.

We propose a machine learning–based framework that builds 3D spatial predictions from a comprehensive stack of environmental datasets available in the United States, including conventional soil maps, climatic and topographic covariates, and point data. We use this framework, visualized in Fig. 1, to provide full-coverage gridded information for the conterminous United States, which can be rapidly updated as revised observations of soil properties and classes become available. The soil property maps we produced include bulk density, pH, percent sand, clay, percent organic C, and total N at seven standard soil depths (0, 5, 15, 30, 60, 100, and 200 cm). Soil class maps include US Soil Taxonomy great group (GG) and modified particle size classes (mPSCs). Modified particle size class is a functional descriptor in US Soil Taxonomy designed to combine agricultural and engineering soil particle size classification systems (Soil Survey Staff, 2010) and has been used to identify areas of similar ecological potential to aid in land management (Nauman and Duniway, 2016). We provide model fitting and accuracy assessment results and discuss advantages and disadvantages of our approach. All soil maps described in this paper are available under the Open Database License v1.0 and can be obtained from the Penn State University repository at https://scholarsphere.psu.edu.

## MATERIALS AND METHODS
### Spatial Prediction with Machine Learning

For this study, we used two types of models for generating predictions: (i) For soil class–type variables (GG and mPSC), we used the random forest method (Breiman, 2001), as implemented in the ranger package (Wright and Ziegler, 2017) for R (R Development Core Team, 2009), and (ii) for soil property numeric–type variables, we used an ensemble of two tree-based machine learning methods—random forest (Breiman, 2001) and gradient boosting (Hastie et al., 2009)—as implemented via the ranger package (Wright and Ziegler, 2017) and the xgboost package (Chen and Guestrin, 2016), which are both available as R contributed packages (R Development Core Team, 2009).

Random forest is a nonparametric model based on pattern recognition using similarities among observations to fit decision trees. To build a tree, a bootstrap sample (a random sample with replacement) is taken from observations. To determine a split at a node in a tree, a random subsample of predictor variables is taken to select the predictor that minimizes the regression error. The size of the predictor subsample is usually the square root of the total number of predictor variables. Nodes continue to be split until no further improvement in error is achieved. Omitted observations, termed the "out-of-bag" sample, are used to compute the regression errors for trees. Gradient boosting builds additive regression models by sequentially fitting a parameterized function or base learner to current "pseudo" residuals by least squares at each iteration. The pseudo-residuals are the gradient of the current loss function being minimized at each iterative step (Hastie et al., 2009).

For soil class–type variables, we used a general spatial prediction model of the form:

$$\text{soil class probabilities }(s) = f[X_1(s) + X_2(s) + X_3(s) + \ldots + X_p(s)]$$

where $s$ is two-dimensional space (northing, easting), and $X$ values are covariates derived from the stack of environmental datasets, which were either aggregated from finer-resolution (e.g., 30 m) data or downscaled (using cubic spline or averaged) from coarser (e.g., 1 km or 250 m) resolution data to 100 m.

For soil property numeric–type (3D) variables, the model was:

$$\text{soil property }(s,d) = f[d + X_1(s,d) + X_2(s,d) + X_3(s,d) + \ldots + X_p(s,d)]$$

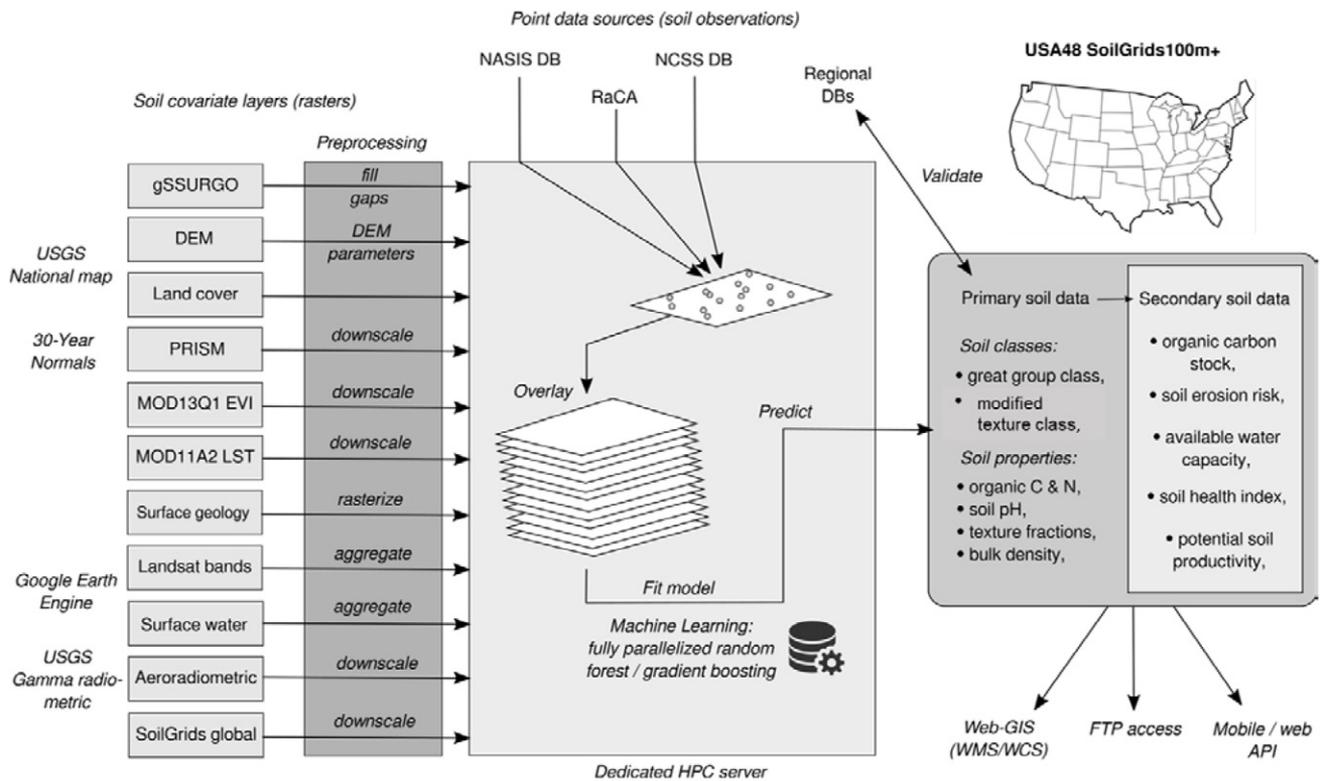

**Fig. 1. General data processing workflow used to generate spatial predictions at 100 m resolution (USA48 SoilGrids100m+).**

where $s$ is two-dimensional space (northing, easting), $d$ is depth, and $X$ values are covariates, including global predictions of soil properties (SoilGrids), which were downscaled from 250-m to 100-m resolution using cubic spline interpolation.

This method of integrating global soil model predictions into our model formulation can be considered an ensemble prediction or model averaging, operating under the assumption that combining realizations of target variables is better than one realization alone. Mulder et al. (2016) correctly recognized that, in many areas in the world, locally produced predictions of soil properties would likely be significantly more accurate than SoilGrids. Because each contributor model has its own strength and weaknesses, we decided to combine global and local predictions rather than selecting the single best-performing model for a given situation or scenario (the traditional pursuit). A caveat of this modeling decision is that predictions can be vulnerable to error propagation, resulting in optimistic cross validation results. If available, the use of an independent testing dataset for validation can produce more robust validation metrics, or an empirical uncertainty function can be used to scale prediction probabilities for class models, as was done in this study. No model variograms for residuals or kriging of predictions with residuals were done because some research has shown this to be unnecessary for variables where covariates explain >60% of spatial variation (Lacoste et al., 2016; Vaysse and Lagacherie, 2015).

The ranger and xgboost packages were selected for predictive modeling because (i) they have high prediction accuracies when modeling complex interactions among predictor variables (Cutler et al., 2007; Hengl et al., 2017) and (ii) they can be implemented in parallel, so that predictions can be generated for large volumes of pixels.

The number of trees grown in the random forest model was reduced from the default setting of 500 to minimize the risk of model overfitting. This was also done because previous research showed no significant effect on model accuracy from reducing the number of trees in the model (Latinne et al., 2001; Oshiro et al., 2012). After initial testing, the number of trees grown for all models was set to 85. The train function of the caret package (Kuhn and Johnson, 2013) was used to tune model parameters, and final parameter selection was based on reducing the RMSE over 100 iterations. The parameters tuned included eta and max_depth, and final values were 0.4 and 3 for soil property models. Eta controls the learning rate of the forest by scaling the contribution of each tree between 0 and 1, whereas Max_depth is the maximum depth of a tree. For soil property models, the xgboost tuning parameters γ, colsample_bytree, and min_child_weight were held constant at 0, 0.8, and 1, respectively. Gamma is the minimum loss reduction required to make a partition on a leaf node of a tree, Colsample_bytree is the fraction of columns to sample from when constructing a tree, and min_child_weight is the minimum number of instances set in a terminal node (Chen and Guestrin, 2016). Tuning results of mtry (the number of variables randomly sampled at each split in the random forest model) are reported in Supplementary Table S1.

The general spatial prediction framework, described in detail in Fig. 1, includes a number of GIS operations, such as downscaling, aggregation of rasters, spatial overlay, fitting of machine learning models, generation of predictions, and data export. All processing was implemented on a 508-RAM high-performance server with 48 cores and 4 gigabytes of disk space. For soil properties, model fitting and spatial prediction (values at seven standards depths) took approximately 12 h per property to build models and predict; soil classes took approximately 14 h to build models and generate spatial predictions.

Final soil properties modeled and mapped included (i) soil organic C in % weight, sand and clay in % weight, bulk density of the fine earth fraction (<2 mm) in $g\,cm^{-3}$, total N in % weight, and soil pH in 1:1 soil–$H_2O$ solution. The USDA taxonomic soil classes modeled and mapped included mPSCs (78 classes) and soil GG (291 classes).

### Data

For model training we used point data from three data sources: (i) The National Cooperative Soil Survey (NCSS) Characterization Database, (ii) The National Soil Information System (NASIS), and (iii) The Rapid Carbon Assessment (RaCA) Project.

### National Cooperative Soil Survey Characterization Database

The NCSS characterization database provides soil characterization data from projects performed by the Kellogg Soil Survey Laboratory and cooperating laboratories (http://ncsslabdatamart.sc.egov.usda.gov/). Beginning with the 2015 NCSS database, 34,183 pedons (213,499 horizons) were available for spatial modeling. This database was filtered to retain only soil samples that had a minimum of a site location (WGS84 coordinates) and at least one soil property observation.

The degree of completeness for each soil pedon and the available property and attribute data for each individual sample varied cross the database. Most data had soil organic C, soil texture (total sand, silt, clay content), and pH. For the organic C data, a linear regression based on Wills et al. (2013) was applied to Walkley–Black organic C measurements to align with total combustion C measurements. Data preprocessing was done to remove negative values for soil properties. Compiled soil sample data contain variations that can come from several potential error types in addition to the actual property variations. These include the level of accuracy recorded, measurement error, the date at which the sample was taken, and imprecise methods of locating sampling locations. These factors can introduce biases into the models because these differences may be confounded with changes in property or class values. Corrections can be made for some factors to reduce this possibility, as was done for organic C, but overall there is limited ability to isolate or negate the effects of these factors on derived models.

### National Soil Information System Data

The USDA–NRCS maintains a large database of soil maps and observations called NASIS. We used NASIS records made primarily during soil survey activities (separate from NCSS characterization data). These observations are mostly field observations made by scientists using transects to determine SSURGO components and to build map units. The NASIS point data are not given the same review process as the component/map unit data (SSURGO). Most have not been tested in a laboratory but were described and classified to SSURGO components and US Soil Taxonomy classes (Soil Survey Staff, 1999, 2014) by experienced local soil scientists.

The most recent taxonomic classification for each observation was queried from NASIS using R (R Development Core Team, 2009). Soil great group (GG; e.g., "Calciargids") and particle size classes (PSCs) (e.g., "Loamy" or "Coarse-loamy") were then extracted from taxonomic class names. Particle size classes are one of the most consistently defined taxonomic descriptors in US Soil Taxonomy (Caudle et al., 2013; Duniway et al., 2010). These classes generalize the texture of a soil profile based on a representative set of depths called the control section, which is specific to different groups of soils (Soil Survey Staff, 2014). Extracted PSCs were then modified (mPSC) as follows to ensure that classes more fully covered the spectrum of possible soils across the conterminous United States (Nauman and Duniway, 2016).

Observations of rock outcrops, which traditionally have no PSC, were labeled as such. To avoid redundancy in the taxonomic classification, Psamments (sandy soils) were assigned to a sandy class because these soils have no PSC designation. Soils with a Lithic (i.e., shallow bedrock contact) taxonomic classification were labeled with a "Lithic" mPSC. Histosols (organic soils), which traditionally have no PSC, were assigned to an "Organic" mPSC. These processing steps resulted in 328,380 GG observations and 308,291 mPSC observations that were used as training data for building soil class models (Fig. 2).

### Rapid C Assessment Dataset

The NRCS RaCA project was conducted to capture information on the C content of soils across the conterminous United States (Soil Survey Staff, 2016). A multilevel stratified random sampling scheme was created to maximize spatial sample coverage. The RaCA project consisted of 31,215 pedon observations with organic C, total N, and bulk density measurements (when possible) in the 0 to 50 profile range. Soil organic C and total N measurements were performed according to standard Kellogg Soil Survey Laboratory methods (Soil Survey Staff, 2014). The RaCA dataset was combined with the NCSS database for predicting soil properties.

### Soil Covariates

We used a range of geospatial datasets relevant to soil formation to build a stack of environmental covariates for predictive soil modeling. Each environmental covariate was sourced at 100-m resolution, or a resampling method was applied to convert data to 100-m resolution (e.g., cubicspline was applied to PRISM datasets to convert from 800 to 100 m). Examples of the environmental covariates are shown in Fig. 3, and a detailed summary of the datasets, sources, and resampling methods is available on Github (https://github.com/aramcharan/US_SoilGrids100m).

The environmental covariates included:

- Digital elevation model (DEM)-derived covariates: Ten terrain attributes were derived from a conterminous 100-m DEM for the United States (http://nationalmap.gov/) using SAGA GIS software (Conrad et al., 2015). These included: elevation, slope gradient, profile curvature, multiresolution index of valley bottom flatness, surface roughness, valley depth, negative topographic openness, positive topographic openness, Melton ruggedness number, SAGA wetness index, and topographic position index.
- PRISM climate covariates: PRISM 30-year normal climate surfaces covering the period 1971 to 2010 included precipitation, mean temperature, minimum temperature, maximum temperature, mean dew point temperature, minimum vapor pressure deficit, and maximum vapor pressure deficit (Daly et al., 2008). Nineteen bioclimatic indices (e.g., annual diurnal range, total precipitation in the driest quarter, average temperature in the driest quarter) derived from PRISM normal datasets by O'Donnel and Ignizio (2012) were included.
- MODIS land products: Satellite imagery datasets derived from 35 yr of MODIS data included six long-term averaged bimonthly mean and SD enhanced vegetation indices, six long-term averaged bimonthly mean surface reflectances for MODIS bands four (NIR) and seven (MIR), 24 long-term averaged monthly mean daytime and nighttime surface temperature measurements, and 12 long-term averaged monthly cloud cover datasets of twice-daily remote sensing–derived cloud observations (https://modis-land.gsfc.nasa.gov).
- Landsat 30-m resolution products: Landsat (cloud-free) NIR, SWIR bands, γ radiometric images (Hansen et al., 2013), long-term surface water occurrence (Pekel et al., 2016), and bare ground images were aggregated from 30 to 100 m using the averaging algorithm in GDAL (Warmerdam, 2008).
- SoilGrids250m: SoilGrids250m (Hengl et al., 2017) soil property maps were included in the covariate stack. The original maps were downscaled to 100 resolution using bicubic resampling in GDAL (Warmerdam, 2008).
- Additional environmental covariates: These included land cover classes for the year 2011 (Homer et al., 2015), US potential natural vegetation (Küchler, 1964), historical natural fire regime classes (Fire Sciences Laboratory, Rocky Mountain Research, 2000), and aeroradiometric grids (Duval et al., 2005).

### gSSURGO Map Unit Data

Parent material classes (87 classes) and aggregated drainage classes (four classes) based on the representative soil components of gSSURGO map units were extracted from gSSURGO. Original parent material classes were summarized to 87 distinct parent material kind classes based on Schoeneberger (2012). These 87 classes were nested into nine parent material kind groups based on the geomorphic description system of Schoeneberger (2012). This relationship is illustrated in Supplementary Table S1. For drainage classes, one limitation of the data was the variation in methodology for mapping drainage classifications across the United States. Given this limitation, the eight drainage classes extracted were reclassified to four drainage classes (Supplementary Table S2) to dampen the effect of inconsistent classification.

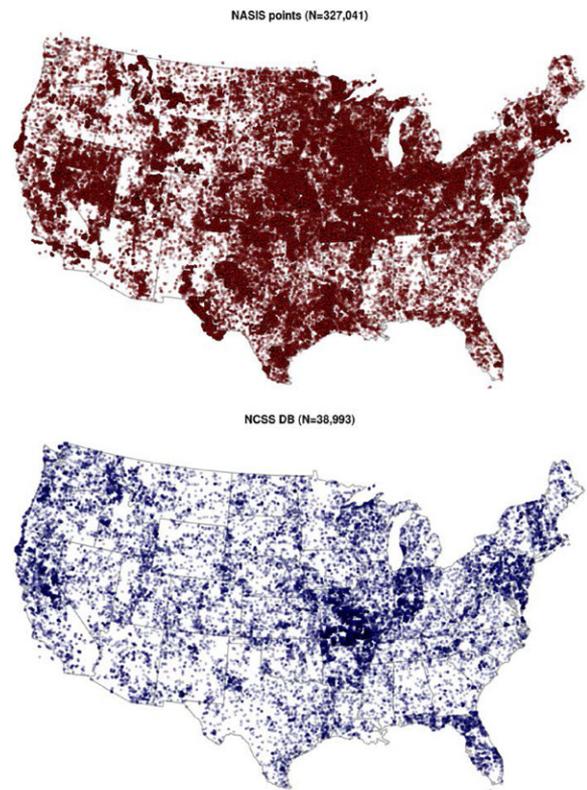

Fig. 2. Maps of points used to train models. Red, NASIS soil class points; blue, NCSS database and RaCA points (combined).

Parent material and drainage class data were missing from approximately 15% of the gSSURGO map units. To use these covariates for spatial prediction (which requires spatially complete covariates), random forest models were built to predict the missing data using the 10 terrain attributes (see Soil Covariates) and the USGS surficial materials map (Soller and Reheis, 2004). Parent material and drainage class prediction accuracies were 54 and 79%, respectively. These models were used to gap fill missing data from SSURGO map units. Random checks were made on predictions before maps were used as model inputs; these results were not independently peer reviewed.

### Accuracy Assessment
#### Cross-validation of Soil Property and Class Models

After reviewing modeling results and completing multiple iterations of the modeling process to eliminate obvious errors (e.g., incorrect trends in results due to political borders), 10-fold leave-location-out cross-validation was run to assess model accuracy for soil property and taxonomic class models. For this validation scheme, soil samples were split into training and testing datasets. SoilGrids250m covariates were removed from the cross-validation to ensure the assessment was completely independent of covariates produced with NCSS pedon data. Model training used 90% of the data, and the remaining 10% were used for model testing; therefore, validation results are more conservative compared with models using 100% of the data to produce maps. Soil samples were split into training and test data such that horizon observations from a single pedon were not split across training and test datasets (leave-location-out scheme). This 90/10 data partitioning scheme was repeated 10 times to calculate the averaged performance metrics. For each of the six numeric soil properties, we derived the $R^2$ value, mean error, mean absolute error, and RMSE (Kuhn and Johnson, 2013). All equations for cross-validation metrics are provided in the supplementary materials.

## Independent Validation and Uncertainty of Soil Class Models

In addition to cross-validation, nine regional datasets were collated from pedon observations and used to independently validate GG and mPSC class predictions. These collated and harmonized validation datasets, summarized in Tables 1 and 2, were collected as part of graduate student theses or as part of NRCS soil survey efforts but had not been included in any of the databases used in model training (Brungard et al., 2015; Fonnesbeck, 2015; Molstad, 2000; Stum et al., 2010). Although coverage of independent data is not exhaustive due to the limited availability of soils data, these regional datasets provide an indication of modeling performance in a range of climates and landscapes across the United States. Validation datasets included data from three areas in western Utah (Beaver County, Juab County, Millard County), southeastern Utah, north-central Wyoming, southcentral New Mexico, northern Minnesota, eastern Iowa, and Maine.

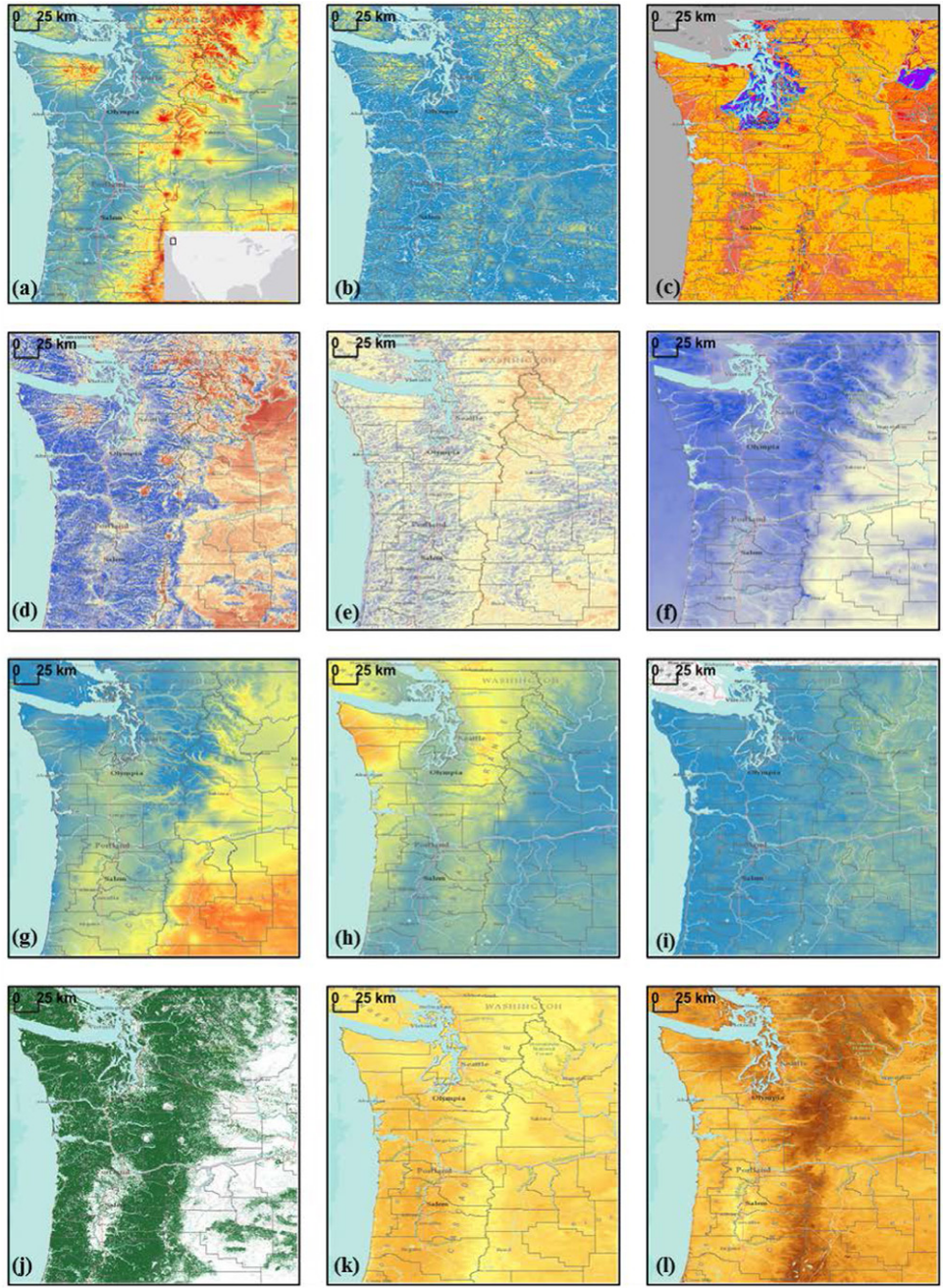

Fig. 3. Examples of covariates used for the conterminous United States. (a) Elevation (http://nationalmap.gov/). (b) Vertical distance to channel network (Conrad et al., 2015). (c) SSURGO parent material (Schoeneberger, 2012). (d) Mean monthly MODIS vegetation index for January and February (https://modis-land.gsfc.nasa.gov). (e) Long-term averaged mean monthly surface temperature (daytime); MODIS, January. (f) Long-term average cloud cover (Wilson and Jetz, 2016). (g) Annual mean diurnal range (O'Donnel and Ignizio, 2012). (h) Total precipitation in the driest month (O'Donnel and Ignizio, 2012). (i) Average vapor pressure deficit (Daly et al., 2008). (j) Tree cover (Hansen et al., 2013). (k) SoilGrids250m Clay (Hengl et al., 2017). (l) SoilGrids250m Sand (Hengl et al., 2017).

Final soil class predictions and the associated prediction probabilities were validated. Overall accuracy was used to validate soil taxonomic class prediction accuracies for the entire validation dataset and for individual validation datasets. Prediction probabilities were validated using empirical uncertainty functions designed to relate the deterministic maps of GG and mPSC to independent field observations. These functions build on methods from several prior soil class mapping studies showing positive but nonlinear trends between rates of independent observation agreement with tree-based machine learning predictions and the underlying prediction probabilities produced by these models (Nauman and Thompson, 2014; Nauman et al., 2014). This process starts by tallying instances where independent field observations are correctly and incorrectly predicted by the hardened deterministic class maps. Then the instances are grouped by intervals of the prediction probabilities from the machine learning model associated with each field observation. The intervals were created to include approximately equal numbers of field observations using the ggplot number function in R (Wickham, 2016). The number of intervals was initially varied in a sensitivity analysis between 10 and 20 to maintain 0.5 to 1% resolution in the estimates of proportions of field observations correctly predicted, heretofore referred to as validation probabilities.

We looked for trends between median prediction probabilities and validation probabilities of the intervals in the different interval binning scenarios. As an example, if the median prediction probability for an interval is low (20), we expect the proportion of independent observations that match this class (i.e., the validation probability) to also be low (and likely lower than the median prediction probability).

Based on the finding that very similar positive trends were found for all numbers of intervals tested, we looked for the best fitting regression model among all interval scenarios using a variety of empirical link functions (including linear, linear-log, quadratic, power, exponential, and generalized additive models) in R. The best models for both mPSC and GG were chosen based on the highest $R^2$ value. The fitted curves were projected onto the mapped prediction probability surface to create a validation probability surface map. This validation probability map estimates how likely it is that the deterministic map is correct at every pixel, which is valuable information for end users. It also provides an idea of overall model uncertainty at a pixel because low deterministic map probabilities indicate more overall model confusion.

## RESULTS
### Model Fitting

Examples of the complete coverage, gridded soil property, and class maps results are illustrated in Fig. 4. Visual analysis of Fig. 5 shows mapped features can be distinguished up to the 1:25,000 scale. In Fig. 4, examples of spatially explicit predictions of percent clay and pH are shown at 5 cm depth. The probability of occurrence of coarse loamy mPSC and Fragiudalfs across the United States is also shown. The probability map of Fragiudalfs aligns with SSURGO, showing that the most extensive areas of occurrence of Alfisols with fragipans is the southern Mississippi valley and the southern Appalachian Mountains (Bockheim and Hartemink, 2013). The probability map of Fragiudalfs in Fig. 5A also shows the probability of occurrence increasing in forested areas, similar to the distribution of fragipan soils reported in Grossman and Carlisle (1969). Figure 5B shows that percent clay at 5 cm depth is lower in areas of high probability of Fragiudalfs (and vice versa), which is in line with findings that fragipans occur primarily in the fine-silty, fine-loamy, and coarse-loamy PSCs (Bockheim and Hartemink, 2013).

In addition to final reported soil property maps, the dataset collection includes preliminary maps of Mg [cmol(+)/kg], K [cmol(+)/kg], and cation exchange activity [cmol(+)/kg] that were initially tested but not validated in this study. Probability maps for each soil GG and mPSC class are also provided. Summary results of the model fitting are shown in Fig. 6 and 7 and Table 3. In Fig. 6, the variable importance (VIMP) indicates the importance of a predictor variable in a model by measuring the increase in prediction error when a predictor variable is "noised up" by randomization (Breiman, 2001). Using variable importance metrics for model interpretation can be limiting where there are groups of highly correlated variables. Strobl et al. (2007) showed that correlated variables are used interchangeably in decision trees of random forest models, resulting in less-relevant variables receiving boosted importance. Therefore, we limited using the results of the VIMP analysis for model interpretation and relation to how soil survey is conducted.

Table 1. Number of modified particle size class observations, great group classes, and observation density in each independent dataset.

| Area location | Number of observations | Number of classes | Observation density (per km$^2$) |
|---|---|---|---|
| Iowa | 69 | 3 | 1.898 |
| Maine | 183 | 14 | 0.001 |
| North-central Wyoming | 51 | 7 | 0.029 |
| Northern Minnesota | 108 | 12 | 0.002 |
| South-central New Mexico | 103 | 8 | 0.042 |
| Southeastern Utah | 356 | 7 | 0.005 |
| Western Utah (Beaver County) | 329 | 13 | 0.04 |
| Western Utah (Juab County) | 208 | 16 | 0.02 |
| Western Utah (Millard County) | 605 | 10 | 0.02 |
| Total | 2012 | | |

Table 2. Number of great group observations, great group classes, and observation density in each independent dataset.

| Area location | Number of observations | Number of classes | Observation density (per km$^2$) |
|---|---|---|---|
| Eastern Iowa | 69 | 4 | 2.531 |
| Maine | 183 | 12 | 0.001 |
| Northcentral Wyoming | 51 | 5 | 0.021 |
| Northern Minnesota | 97 | 13 | 0.002 |
| South-central New Mexico | 103 | 5 | 0.026 |
| Southeastern Utah | 361 | 9 | 0.006 |
| Western Utah (Beaver County) | 326 | 11 | 0.034 |
| Western Utah (Juab County) | 204 | 8 | 0.01 |
| Western Utah (Millard County) | 605 | 12 | 0.024 |
| Total | 1999 | | |

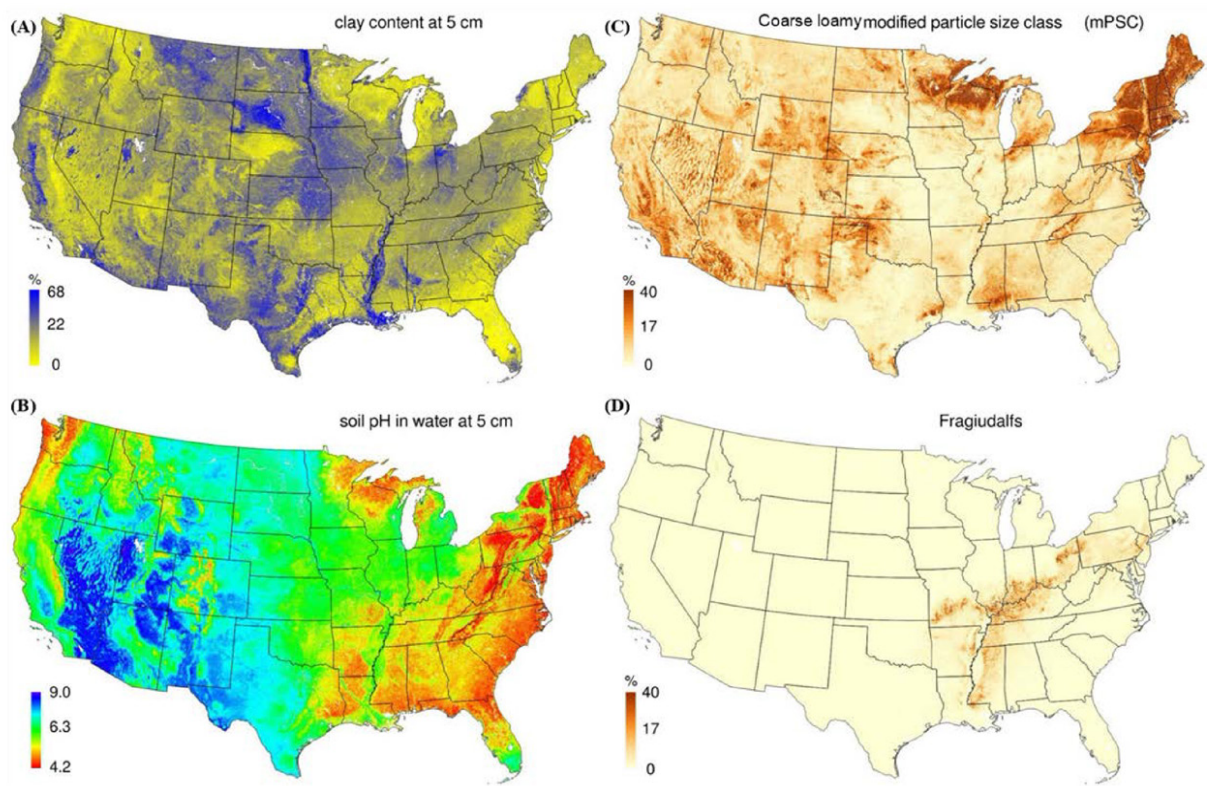

Fig. 4. Examples of soil property and class maps for two soil properties and two soil classes. (A) Percent clay. (B) pH at 5 cm soil depth. (C) Percent probability of coarse loamy modified particle size class. (D) Percent probability of Fragiudalfs great group. The complete dataset collection is available at doi:https://doi.org/10.18113/S1KW2H.

The top 25 predictors reported by the ranger package are shown in Fig. 6. Beyond these 25 top predictors, VIMP was relatively equal among the remaining predictor variables. The drop of importance of covariates is relatively gradual. Depth was a strong predictor across all soil property models, indicating the strong correlation between soil properties and sample depth in the models. Figure 6 also shows that the highly correlated climatic normals and bioclimatic indicators derived from PRISM were the most frequent predictor variables for both soil property and class models. Parameters of the DEM were the second most frequent dataset, followed by SSURGO-derived and satellite data derived from MODIS and Landsat. PRISM data sets, derived from climate-elevation regressions for DEM grid cells (Daly et al., 2008), correlated more strongly overall with soil characteristics than with satellite-derived data, possibly because MODIS/Landsat data reflect dynamic changes in vegetation that may or may not be related to soil properties.

Data from PRISM and DEM parameters, both highly correlated predictor variables, dominated the most important predictors in the soil class models, whereas SSURGO-derived parent material and drainage class predictor variables were present in the top 10 predictors only for soil organic C, total N, sand fraction, and clay fraction. The SSURGO-derived variables appeared not as important as climatic layers or DEM derivatives for mapping GG and mPSC classes. From a practical standpoint, this result may be due to the highly correlated predictor variables dominating the VIMP results or the large number of SSURGO parent material categories used, which resulted in low importance per parent material category. From the perspective of soil survey, GG taxa are largely informed by climate, not by parent material.

## Accuracy Assessment
### Cross-Validation of Soil Property and Class Models

Ten-fold cross-validation was used to calculate the performance metrics of the models for soil properties and classes. We note that, given the spatial clustering of the soil samples (e.g., high sampling density on agricultural land and low sampling density in national forests), it is likely the validation results vary spatially, and we must assume there is spatial autocorrelation of prediction errors. More robust validation results can likely be achieved with a testing dataset collected after a validation strategy, such as an objective probability sampling (Brus et al., 2011). The model with the best performance metrics was soil pH, and the weakest performance metrics resulted from the soil texture models (percent sand and clay). For the soil class models, cross-validation resulted in an average misclassification rate of 0.34 for the mPSC model and 0.40 for the soil GG model, giving mapping accuracies of 66 and 60%, respectively.

We provide mapping accuracy statistics and an empirical uncertainty analysis instead of kappa statistics, which are not as useful for interpretation and comparison of classification accuracy (Pontius and Millones, 2011). Soil property models underestimated values, as indicated by mean error in Table 3. Figure 7 shows the correlation plots for soil properties along with the line of perfect fit (solid line) and the line of best fit (least square method; dashed line). Based on these pH correlation plot, predictions were close to measured values over the range of soil property values. Bulk density, soil organic C (SOC), and total N showed overestimation for low values. There were both over- and under-estimated across the range of values for percent clay and sand, with both models tending more to underpredict high values. Figure 8 shows a comparison of the soil property spatial prediction results with SoilWeb (Beaudette and O'Geen, 2009), the online soil survey for California based on SSURGO and STATSGO for percent clay. Maps derived from completely different modeling approaches show similar spatial patterns.

The resolution of the spatial prediction results is higher (100 m compared with 1 km), but the comparison shows that our results underestimated values in the high range compared with SoilWeb (in line with the results of the performance metrics and correlation plots). Another validation strategy attempted was to extract SOC estimates from the valu1 table associated with the gSSURGOFY2016 product. These were available at standard depth intervals, as opposed to SOC concentrations, which were available at different levels. This comparison seemed less informative because SOC and bulk density are required, and if maps differ it is difficult to determine if it is the SOC or bulk density results that differ from the SSURGO products because SSURGO provides very general guidance to the aggregation of soil properties. Thus, the comparison in Fig. 8 was provided for visualization of patterns or "face validity."

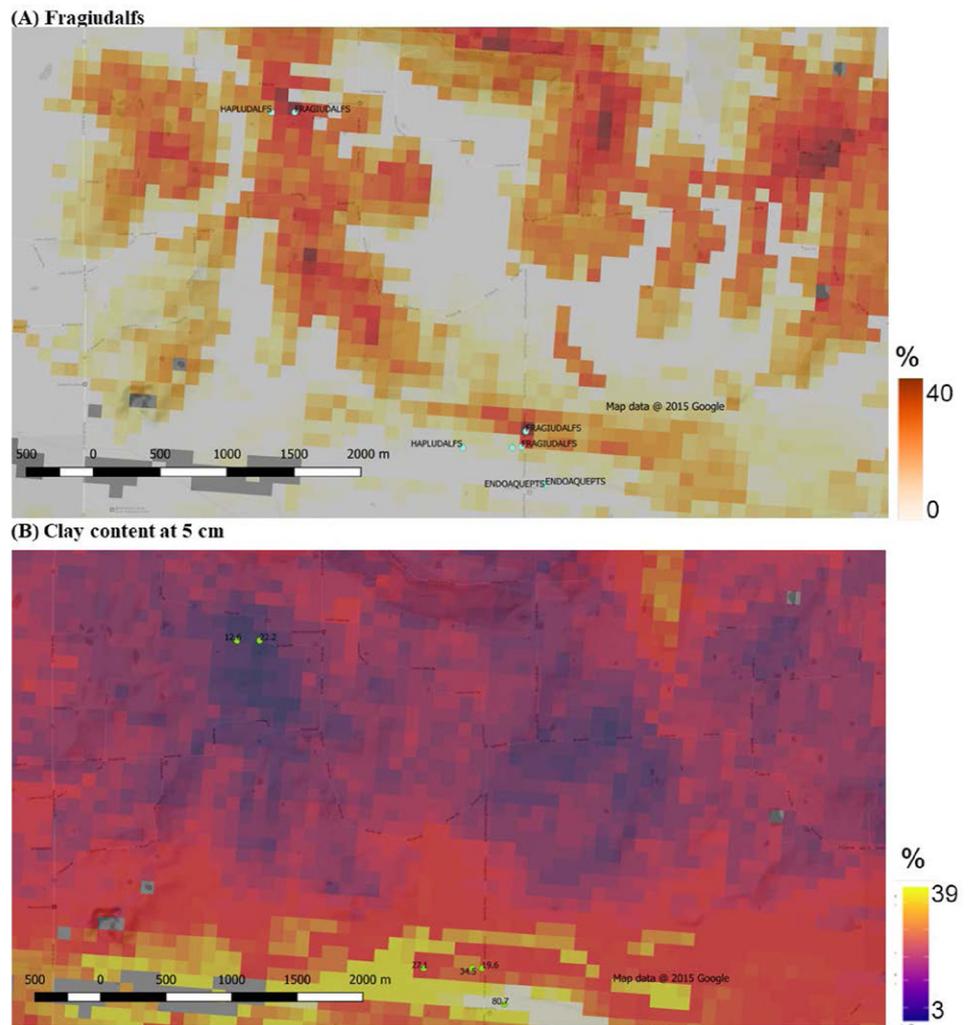

Fig. 5. Zoomed in 1:25,000 scale maps for (A) percent probability of Fragiudalfs and (B) percent clay overlaid on terrain map with training point values displayed. Map location: Little Black Slough Nature Reserve, IL.

### Independent Data Validation of Soil Class Model

The overall accuracy of GG predictions using all independent validation observations was 37%. The overall mPSC accuracy using all independent validation observations was 42%. Validation of GG and mPSC predictions for each independent dataset are presented in Fig. 9 and 10. Great group prediction accuracy ranged between 24 and 58%. Modified particle size class prediction accuracy ranged between 24 and 93%.

### Empirical Uncertainty of Soil Classes

Relationships between the ensemble model prediction probabilities and validation probabilities were quite strong. Linear–log functions with 10 probability intervals fit the GG data best (Fig. 11A), and 12 intervals resulted in the best mPSC fit (Fig. 11B). Strong linear–log relationships for both probability distributions support the hypothesis that the deterministic probabilities produced by random forest models hold valuable information about uncertainty in final predictions but that the relationship to prediction uncertainty is nonlinear.

Overall, mPSC had higher validation probabilities, reflecting the higher overall accuracy statistics. Relatively similar spatial patterns in validation probabilities can be seen by comparing the GG (Fig. 11C) and mPSC (Fig. 11D). Both show better accuracies in the central midwestern states (Nebraska, Michigan, Iowa, Illinois), portions of Texas, and along the eastern coastal plain, likely due to the higher density of soil samples in these areas. The decreased slope of the line at higher prediction probabilities (not actually an asymptote) conceptually represents model overfitting where the prediction probability is higher than the validation probability.

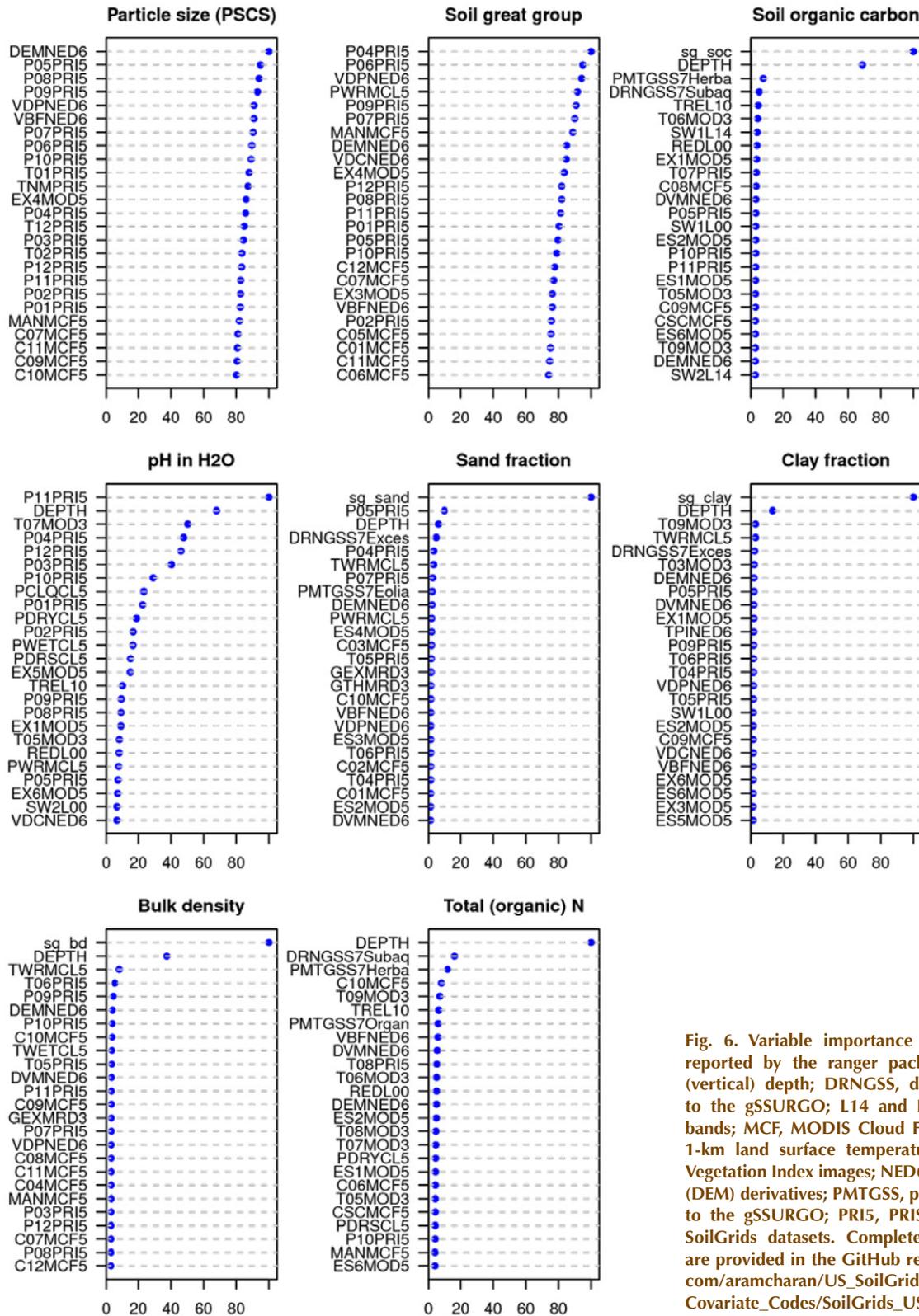

Fig. 6. Variable importance plots per soil variable reported by the ranger package. DEPTH, sampling (vertical) depth; DRNGSS, drainage class according to the gSSURGO; L14 and L00, Landsat cloud-free bands; MCF, MODIS Cloud Fraction images; MOD3, 1-km land surface temperatures; MOD5, Enhanced Vegetation Index images; NED6, digital elevation model (DEM) derivatives; PMTGSS, parent material according to the gSSURGO; PRI5, PRISM climatic images; sg, SoilGrids datasets. Complete descriptions of codes are provided in the GitHub repository (https://github.com/aramcharan/US_SoilGrids100m/blob/master/Covariate_Codes/SoilGrids_USA48_Covs100m.csv).

Table 3. Average prediction error for soil properties based on 10-fold cross-validation.

| Variable | n† | Mean | Min. | Max. | $R^2$ | Mean error | SD | RMSE | Mean absolute error |
|---|---|---|---|---|---|---|---|---|---|
| Soil organic C | 239,526 | 1.62 | 0 | 70.4 | 0.41 | –0.15 | 4.72 | 3.63 | 1.17 |
| % Sand | 195,748 | 34.45 | 0 | 99.62 | 0.57 | –0.37 | 27.1 | 17.8 | 13.5 |
| % Clay | 195,748 | 24.12 | 0.02 | 96.5 | 0.46 | –0.16 | 16.4 | 12.0 | 7.06 |
| Bulk density | 81,541 | 1.38 | 0.06 | 2.6 | 0.42 | 0.01 | 0.27 | 0.20 | 0.15 |
| pH | 200,870 | 6.22 | 2 | 10.9 | 0.68 | 0 | 1.32 | 0.74 | 0.57 |
| Total N | 74,369 | 0.22 | 0.1 | 5.5 | 0.39 | –0.01 | 0.35 | 0.27 | 0.13 |

† Number of samples used for training.

## DISCUSSION
### Machine Learning and Soil Science

This study harnessed the capabilities of a high-performance computing environment along with parallel programmed machine learning algorithms to create soil property and class maps with exhaustive coverage for the conterminous United States (illustrated in Fig. 6 and 7, respectively). We worked within a relatively new scientific paradigm where large datasets from a variety of sources were integrated with quantitative methods to provide gridded spatial variation of soil characteristics. To illustrate our point, we present a SSURGO map unit in Utah where seven soil types are aggregated in a SSURGO association map unit (Fig. 12).

This variation of soil components within map units can limit the usefulness of soils maps for land managers. Previous research shows that this map unit has been implicated as a dust emissions source, possibly causing accidents on Interstate 70 because of heavy grazing and highly erodible soils (Flagg et al., 2014). To manage for that, the variability in soil texture within this map unit is very important. In this case, dominant southwest winds blowing saltating sands from the southwest onto clayey soils derived from Mancos shales is a likely mechanism for dust mobilization (Miller et al., 2012). Therefore, a manager might want to initially focus on stabilizing the sandier areas upwind so they do not bombard the finer areas with saltating sands. However, this solution might not occur to a manager without the spatial context these maps provide. This example demonstrates a worst-case scenario in areas of the United States where detailed soil mapping is limited.

Our results were made possible through collaboration with soil and computer scientists, ensuring expert knowledge was captured and used to validate results. The collaboration with soil scientists within government agencies and land grant universities enabled us to make many decisions on how to prepare the data, run the analyses, evaluate errors and inconsistencies in the map products, conduct visual and quantitative validations of results, and inform future research efforts within this domain. With awareness of the "data deluge" in many sciences (Roudier et al., 2015), we developed our analysis in an open science framework, providing the necessary metadata, data, and code in an open-access Github repository (https://github.com/aramcharan/US_SoilGrids100m).

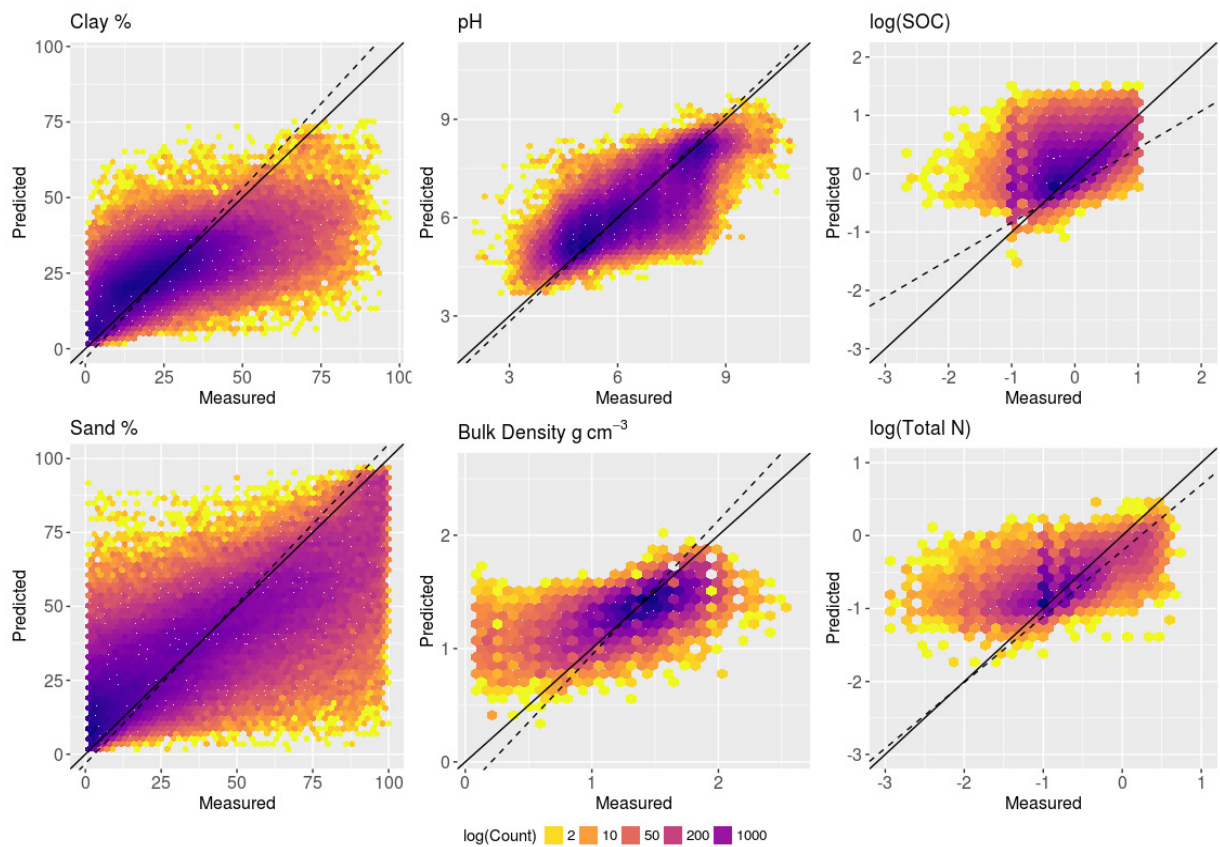

Fig. 7. Correlation plots, based on cross-validation results, for each soil property with the line of perfect fit (solid line) and the line of best fit (least squared method) (dashed line).

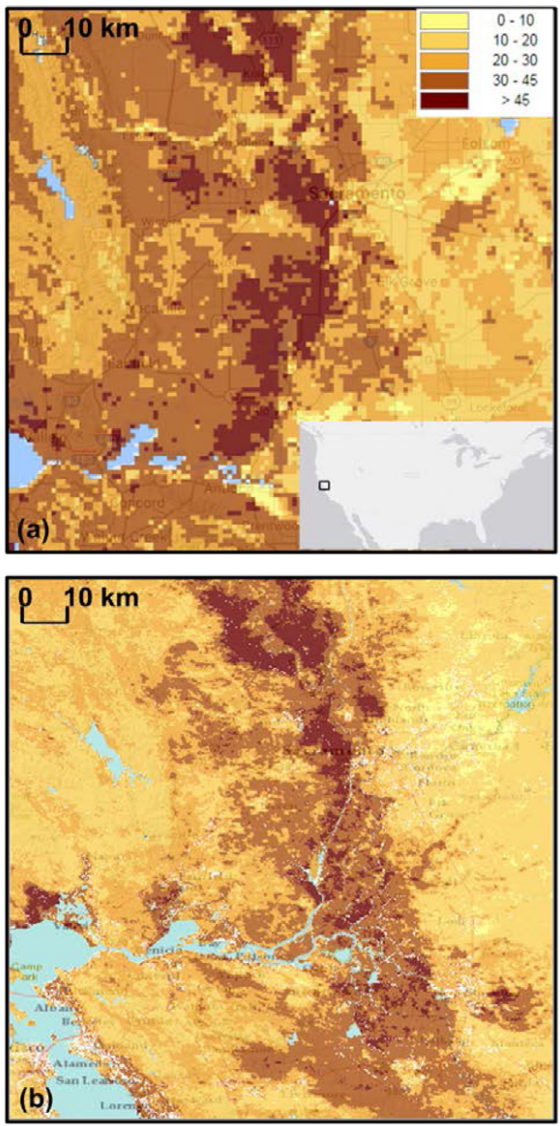

**Fig. 8. Comparison of SoilWeb Soil Properties (Beaudette and O'Geen, 2009) (a) near Sacramento, CA (inset of a) with US48 SoilGrids100m+ results and (b) for percent clay. Legends are the same for both data sources. SoilWeb properties are derived from SSURGO and STATSGO.**

## Limitations of Working with Conventional Soil Data

Our work attempted to address some of the main challenges of working with conventional soils datasets. These include working with map units comprising multiple soil components, managing missing data, identifying taxonomic classifications from multiple editions of US Soil Taxonomy, and assessing the accuracy of soil maps when independent data are limited. Our approach to these challenges is described below.

Currently, the SSURGO database represents soil variability at scales within individual map unit polygons by assigning composition (or proportion) of multiple SSURGO components as well as inclusions of minor soil components and nonsoil areas to map unit identifiers (Nauman and Thompson, 2014). Previous research has worked to disaggregate map units with soil–landscape models (Bui, 2003; de Bruin et al., 1999). Chaney et al. (2016) provide an example of this for the conterminous United States. To exploit information represented in the SSURGO database, we applied previous research that assembled soil parent material classes, based on the representative component percentage for SSURGO, to create a parent material conterminous map (https://gdg.sc.egov.usda.gov/). The tradeoff in selecting a representative soil component for a map unit was the loss of information contained within the map unit pertaining to additional soil components. We also did not have the actual location of components within the map unit, which would have a varying effect on models because the number of components changed and the size of the map unit changed.

Another limitation of working with conventional soil data was the classification of GG data using multiple editions of US Soil Taxonomy. Of the 328,380 observations of most recent GG extracted from the NASIS database, 6622 had obsolete classifications according to the 12th edition of US Soil Taxonomy (Soil Survey Staff, 2014). To make the most of conventional soil data, future work could include recorrelating obsolete data. With multiple editions of GG classes in the training data, there is the potential for confusion between overlapping taxonomic concepts, which could reduce model accuracy. If model accuracy is high, the confusion matrix results for the GG model could inform strategies to recorrelate obsolete classes by providing insight into which GG classes misclassify with each other.

Overall accuracy metrics for the soil taxonomic class models were acceptable considering the geographic extent of the model. However, differences in validation accuracy between independent validation datasets indicate spatially variable prediction accuracy. The lack of a clear trend between the number of observations, the number of classes, observation density (Tables and 2 and 3), and prediction accuracy (Fig. 9 and 10) indicates that differences in validation accuracy between areas is not because of differences in the independent validation data. It is difficult to specify which conditions caused the variability in accuracy metrics between independent validation datasets. However, it is likely that this variability is a result of several factors, including the frequency distribution of observed soil taxonomic classes (Brungard et al., 2015) and similarity between taxonomic classes (Rossiter et al., 2017). Differences in prediction accuracy between areas are also likely a result of differences in the strength of covariate–class relationships.

Further improvement in predictive accuracy results is needed and will likely come from model stratification, the addition of regionally important covariates (e.g., soil erosion and disturbance rates) and management practices (e.g., irrigation), and the inclusion of additional training observations at locations of low prediction quality. Variability in accuracy metrics (and thus variability in prediction quality) likely also results from a mismatch between grid size predictions (100 m) and the inherent scale at which soil characteristics vary across different landscapes. In our visual review of the predictions (data not shown), it appeared that for some landscapes 100 m was too coarse to capture the geographic scale of soil GG taxonomic variability. Using eastern Iowa as an example, the number of GGs occurring at fine scales depends on the landscape and the proximity of soil features to taxonomic breaks. In this case, there are three geologic strata exposed through erosion with different mineralogies and erosion histories. These pedons also straddle taxonomic breaks with diagnostic horizons and soil features found in multiple taxa without large differences in properties. This highlights the difficulties in predicting taxonomic classes when the similarities between taxa are not consistent between higher taxa and across soil landscapes. It is possible that soil taxonomic class predictions could improve if environmental covariates at resolutions <100 m were used for modeling (e.g., Nauman and Duniway, 2016). However, this would increase computation complexity for a model of this large an extent and limit the number of usable covariates (e.g., PRISM data should not be downscaled to 30 m where microclimate effects increase in importance).

Empirical uncertainty functions were created for soil class maps to compare deterministic maps of GG and mPSC with the independent validation data. The empirical uncertainty function predicted validation probabilities that represent estimates of the independent validation data match rates that are more realistic representations of local model uncertainty at each pixel. With the high $R^2$ values for the models to rescale prediction probabilities (0.84 and 0.91 for GG and mPSC, respectively), the uncertainty analyses suggest the probabilities produced by the models are informative even though they were lower than expected.

The disadvantage of these analyses is that the deterministic maps used to validate independent observations only use the top predicted class (Top-1 prediction), reducing the output of the models from a distribution of soil class probabilities to one predicted class. For soil class data with many classes and/or class boundaries that are abstract or overlap conceptually, there would be a considerable loss of information and a perceived low model performance. This is evident in comparing the results for the validation probabilities for the two soil classes mapped. With more soil GG classes (291 classes), the validation probabilities were lower than the mPSC classes (78 classes). Future work can repeat the uncertainty analyses with the top five predicted classes, as is often reported in the machine learning literature (Szegedy et al., 2016). Alternatively, we can compare results with another product at similar scale (gSSURGO or STATSGO).

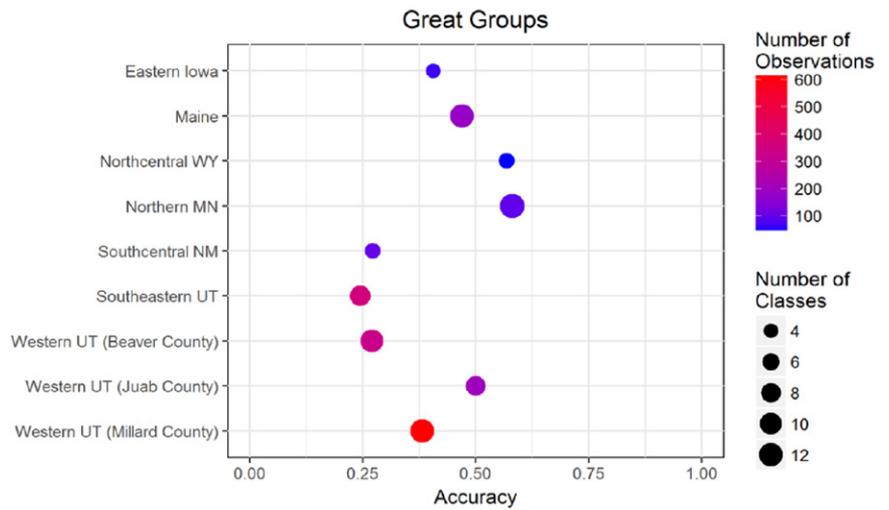

Fig. 9. Great Group prediction accuracy by independent validation dataset. Number of observations is the number of great group observations in each validation dataset. Number of classes is the number of great group classes in each validation dataset.

The results of the soil class models demonstrate the challenges of working with conventional soil concepts. Our results show that there are categories within US Soil Taxonomy that can take advantage of soil mapping concepts without much revision by soil science experts (mPSC), whereas other categories require extensive input by experts to update conventional soil data to apply these methods (GG). Modifications of USDA PSCs similar to what we mapped (mPSC) have proved to be useful for regional-scale land management (Nauman and Duniway, 2016; Nauman et al., 2017), and our results support further development.

## CONCLUSION

This work demonstrated the use of a high-performance computing environment and ensemble machine learning methods to produce gridded soil property and class maps for the conterminous United States at 100 m. The results of this study provide documented estimates and uncertainties of machine learning predictions, given current available national soils and environmental datasets, of six soil properties (percent organic C, total N, bulk density, pH, percent sand, and clay) at seven standard depths (0, 5, 15, 30, 60, 100, and 200 cm) and two soil classification schemes (GG and mPSC). By implementing the study within an automated framework, gridded maps with exhaustive coverage can be rapidly improved (<1 wk of computing) and shared as more data become available and more users test these data products.

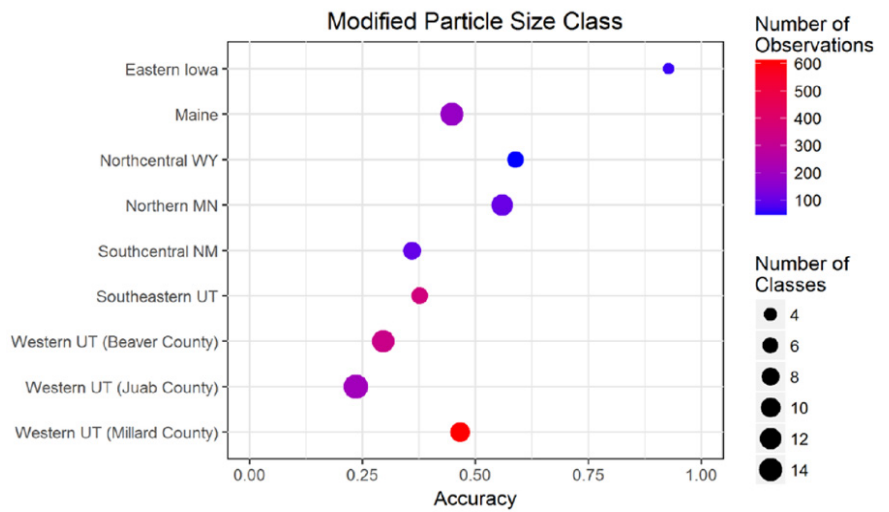

Fig. 10. Modified particle size class (mPSC) prediction accuracy by independent validation dataset. Number of observations is the number of mPSC observations in each validation dataset. Number of classes is the number of mPSCs in each validation dataset.

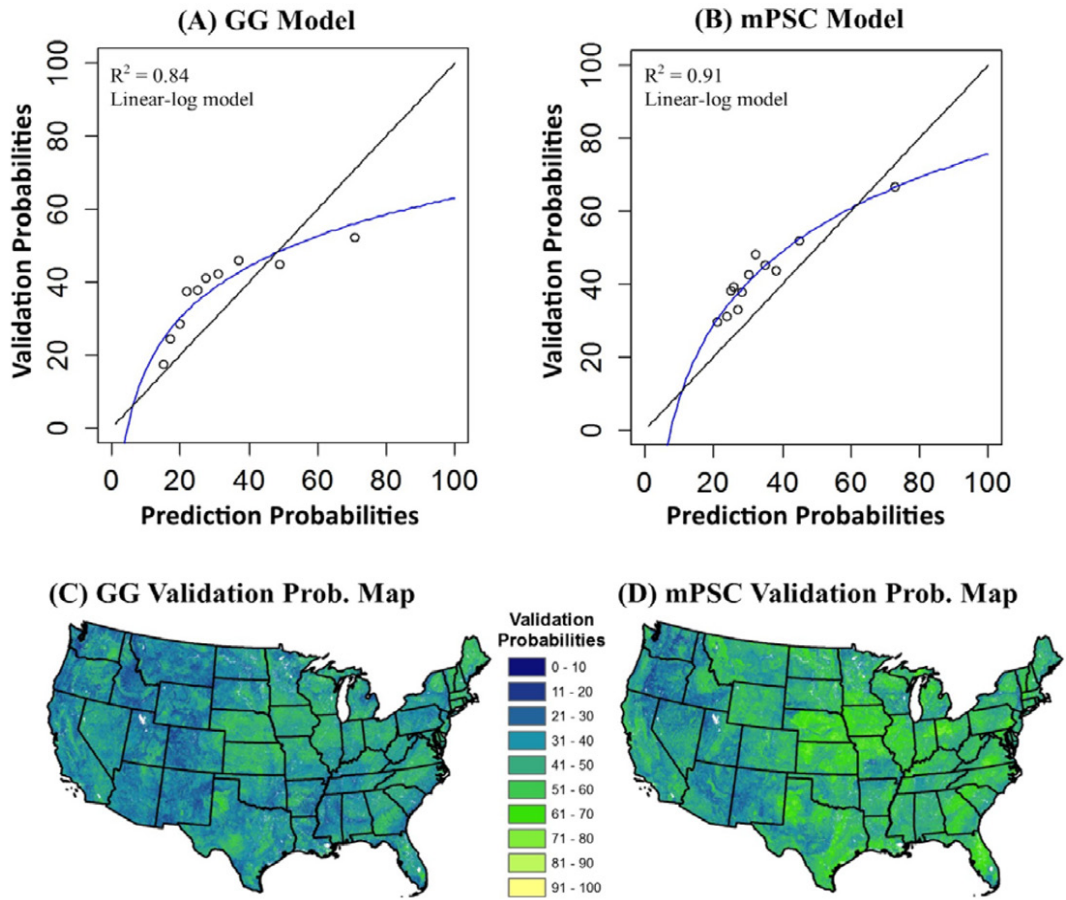

Fig. 11. Empirical uncertainty analyses results. (A) Empirical relationships between prediction and validation probabilities for great groups and (B) modified particle size classes. (C) Resulting maps for great groups and (D) modified particle size classes. Graphs (A) and (B) include a 1:1 reference line (black). GG, great group; mPSC, modified particle size class.

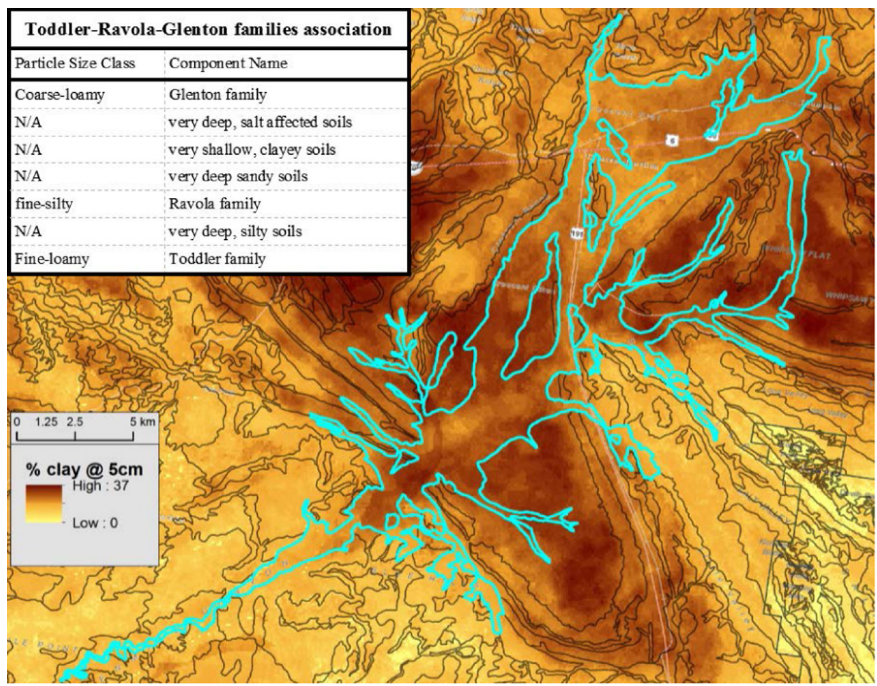

Fig. 12. The Ten-Mile Canyon region near Moab, UT, was mapped with large association map units that often aggregate soils with widely varying management strategies associated with dust emission. This map shows that in terms of near surface clay content variation and variable attribution of SSURGO map unit component.

The average $R^2$ for soil property models was 0.49, and soil class models had an average classification accuracy of 63%. Independent data were used to validate soil class model predictions (1999 points for GG and 2012 points for mPSC) and to conduct empirical uncertainty analyses. Nine regional datasets were used to validate soil class maps, and, al-though an evaluation was not conducted for the conterminous United States, we consider the test case areas a step forward in the process.

This collaborative study attempted to extend the current soils data resources for the United States to meet the needs of geographic simulation models dependent on spatially explicit, 3D reference soil geographic datasets by creating a product that is easier to integrate with models compared with multicomponent map units. Even though mapping accuracy is variable and likely lower than gSSURGO in some areas, the soil property maps can immediately be used for refinement/testing of human landscape models and can be used by land managers who require spatial variability of soil characteristics within SSURGO map units.

The data, metadata, covariates, models, and codes can be freely accessed via doi:https://doi.org/10.18113/S1KW2H and from a public Github repository (https://github.com/aramcharan/US_SoilGrids100m).

## SUPPLEMENTAL MATERIAL

Supplemental information provides data on parent material and drainage classes, model tuning parameters, and cross validation equations.

## ACKNOWLEDGMENTS

This project was supported by the Agriculture and Food Research Initiative Competitive Grant 2012-68005-19703 from the USDA National Institute of Food and Agriculture. The authors thank the reviewers for their edits to this manuscript; the USDA–NRCS (specifically the US National Cooperative Soil Survey Soil Characterization database and the National Soil Information System), USGS, and NASA for creating the national datasets and providing numerous GIS layers that were used as input to modeling; the open source software developers, especially the creators of the packages ranger (Wright and Ziegler, 2017), xgboost (Chen and Guestrin, 2016), and caret (Kuhn and Johnson, 2013); and the creators and maintainers of GDAL (Warmerdam, 2008) and SAGA GIS (Conrad et al., 2015). These maps were made possible thanks to a generation of soil surveyors that have invested 100+ years to collect all ground observations of soil properties and systematically document them.